\documentclass[11pt]{JHEP3}
\usepackage{latexsym}
\usepackage{graphicx}

\def\bea{\begin{eqnarray}}
\def\eea{\end{eqnarray}}
\def\be{\begin{equation}}
\def\ee{\end{equation}}
\def\nn{\nonumber}
\def\a{& \hspace{-7pt}}

\def\Z{{\bf Z}}
\def\al{\alpha^\prime}
\def\hs{\hspace{10pt}}
\title{String vacua with flux from freely-acting obifolds }
\author{M.~Serone and M.~Trapletti\thanks{From November 2003 at the
Deutsches Elektronen-Synchrotron (DESY), Theory Group, Hamburg.}\\
International School for Advanced Studies (SISSA/ISAS) 
and INFN, Trieste.\\
E-Mail: \email{serone@sissa.it}, \email{traplett@he.sissa.it}}

\abstract{
A precise correspondence between freely-acting orbifolds (Scherk-Schwarz
compactifications) and string vacua with NSNS flux turned on is established
using T-duality.

We focus our attention to a certain non-compact $\Z_2$ heterotic
freely-acting orbifold with ${\cal N}=2$ supersymmetry (SUSY).
The geometric properties of the T-dual background are studied.
As expected, the space is non-K\"ahler with the most generic torsion
compatible with SUSY.
All equations of motion are satisfied, except the Bianchi identity for the
NSNS field,
that is satisfied only at leading order in derivatives, {\em i.e.}
without the curvature term. We point out that this is due to unknown corrections
to the standard heterotic T-duality rules.
}
\preprint{SISSA-88/2003/EP}


\begin{document}
\section{Introduction}

String vacua with fluxes have recently received a revival
of interest, mainly due to the fact that they typically lead
to a less number of unwanted string moduli.
Examples of string/M-theory models in presence of background fluxes
have been constructed in \cite{BB1,drs}, and (more recently) in \cite{K1,K2} and
\cite{B1} in the context respectively of IIA/B or heterotic/M-theory.
It turned out that there is a relation between certain flux compactifications
with Scherk-Schwarz (SS) compactifications \cite{SS,SSstringaH}
(see {\em e.g.} \cite{K3,Dab}\footnote{Such
a relation can also be analyzed from the low-energy effective action point of view,
where in both cases gauged supergravities arise \cite{gauged}.}),
or equivalently twisted tori \cite{melvin,Russo,Dab}, {\em i.e.}
tori with non-trivial periodicity conditions along its cycles.
In the context of string theory, SS compactifications are conveniently
described in terms of freely-acting orbifolds (see \cite{SSstringaH-free} and
\cite{SSstringaO} for explicit 4D constructions, respectively in heterotic
and IIB orientifold models).

Crucial in establishing a relation between string vacua with flux and
SS compactifications is T-duality, that essentially maps a non-trivial twist to a
non-vanishing flux. Although the basic idea leading to the above relation is straightforward
and clear, so far there has been not even a single example of how this correspondence
is implemented in a consistent string set-up.

Aim of this paper is to explicitly show how a particular class of string models
with SS compactification is related by T-duality to backgrounds with flux,
leading then to a precise link between the two constructions.
We focus our attention to an heterotic model, with ${\cal N}=2$ space-time supersymmetry.
The model is based on a simple $\Z_2$ freely-acting orbifold, where the $\Z_2$ acts
at the same time as a reflection in four coordinates and a shift in an other coordinate.
It corresponds to a SS compactification, where the
twist is geometric, being a subgroup of the local Lorentz group in the internal directions.
These vacua have also an alternative description as generalized Melvin backgrounds \cite{melvin}.
In order to perform a T-duality transformation on these backgrounds, it is necessary
to identify an isometry of the background, isometry that turns out to be present
only in the large-volume limit for the 4 directions reflected by the $\Z_2$ orbifold element.
In this non-compact limit, we obtain by T-duality an  heterotic model
with a non-vanishing Neveu-Schwarz/Neveu-Schwarz (NSNS) 3-form field strength turned on.
A similar (actually simpler) procedure can be performed to Type IIA or Type IIB
string models to get vacua with non-vanishing NSNS flux.

Besides establishing an explicit and consistent duality between string vacua
on twisted tori and fluxes turned on, the above construction is also a useful tool
to build interesting new string vacua.
The geometric properties of our heterotic background with flux, in particular the torsion classes,
are studied, along the lines of \cite{Gurrieri2,Martelli}.
It is found that
${\cal W}_3$, ${\cal W}_4$ and ${\cal W}_5$ (see \cite{DA1} for the notation and further details)
are all non-vanishing and satisfy the constraints imposed by SUSY, namely
$2{\cal W}_4+{\cal W}_5=0$, with both ${\cal W}_4$ and ${\cal W}_5$ real and exact.
All the supersymmetric conditions obtained by \cite{Candelas,Strominger,Gates,deWit}
are satisfied exactly by
our background. On the contrary, the Bianchi identity for $H$ is satisfied only
at leading order in derivatives, {\em i.e.} without the curvature correction.
This is due to the T-duality rules we are using \cite{Hassan1,Hassan2} that are valid only
at leading order in a derivative expansion, and hence miss the $R^2$ curvature
correction\footnote{We thank S.F. Hassan for pointing this out to us.}.
It is in fact known that the usual T-duality rules for curved backgrounds
\cite{Buscher} in general get higher-order corrections (see {\em e.g.} \cite{Kal}).

The above construction represents one of the few
possibilities to build explicit and calculable
heterotic or Type IIB string backgrounds with non-vanishing flux.
In the latter case, by means of a chain of S- and T-dualities, one can also
build backgrounds with NSNS and Ramond-Ramond (RR) fluxes turned on.
Unfortunately, our procedure is valid only for non-compact string backgrounds
and thus at this stage does not allow to construct new compactifications down to
four dimensions.
Nevertheless we hope that this technical problem might be
solved, opening in this way the possibility of studying semi-realistic
string backgrounds with non-vanishing flux at a full-fledged string level.
In order to satisfy the Bianchi identity for $H$ and get a full consistent set-up in this way,
it is important
to better study the T-dualities rules for heterotic curved backgrounds
that do not seem to be completely established so far.

The paper is organized as follows. In section 2 we show how to construct a SS heterotic
string compactification from a freely-acting orbifold, and how it is related to
a generalized Melvin background. In section 3 the T-dual model with fluxes is obtained
and its geometric properties briefly analyzed. In an appendix some details on the
background with flux are reported.

\section{Freely acting orbifolds and generalized Melvin backgrounds}

Let us consider the heterotic string ($SO(32)$ or $E_8\times E_8$)
on $R^4\times S^1\times (T^4\times S^1)/{\bf Z}_2$, where the $\Z_2$ acts
as $z_{k}\rightarrow \exp(2 i\pi v_k) z_k$ on the two complex coordinates
$z_{1,2}$ of $T^4=T^2\times T^2$, where $v_k = (-1/2,1/2)$, and at the same
time as an half-shift on the circle: $x\rightarrow x+\pi R$, where $R$ is the radius
of the circle. For simplicity, we take the two tori to be rectangular,
with complex structures $\tau_n=i$, $n=1,2$.

This model is a simple supersymmetric freely-acting compact orbifold model.
It corresponds to a Scherk-Schwarz compactification, where fields
are twisted according to their $SO(4)$ Lorentz quantum numbers along the direction
of $T^4$. In 4D notation, the model has ${\cal N}=2$ supersymmetry.
Modular invariance imposes that a rotation has to be implemented
also in the internal lattice part. In a fermionic world-sheet formulation in terms
of 16 complex fields $\lambda^A$, $\Z_2$ acts as $\lambda^A\rightarrow \exp(2i \pi V_A)
\lambda^A$, $A=1,\ldots,16$. Modular invariance imposes that
$v_k^2 - V_A^2 =n$, with $n$ any integer number. For simplicity, we focus
in the following to a standard embedding, in which $V_A=(1/2,-1/2,0,\ldots,0)$.

The 1+1 dimensional $\sigma$-model associated to this model is an exact
free super conformal field theory (SCFT) (in the RNS formalism), with the above
identification for the fields. In an ${\cal N}=1/2$ superspace language \cite{Hull:jv},
the relevant SCFT associated to the 5 directions $(z_1^{(0)},z_2^{(0)},x)$ and
the $\lambda^{A(0)}$'s fields is the following:
\be
{\cal L} = -i\int\! d\theta \Bigg[D X \bar \partial X +\frac 12 \sum_{k=1,2}
(D Z_k^{(0)} \bar\partial \bar Z_k^{(0)} + D \bar Z_k^{(0)} \bar\partial Z_k^{(0)})
-i \sum_{A=1}^{16} \Lambda_A^{\star (0)}D \Lambda^{A(0)} \Bigg]\,,
\ee
where $X,Z_k^{(0)}$ are superfields of the form
$\Phi^k  =  x^k + \theta \psi^k$, with $\psi^k$ left-moving world-sheet fermions,
and $\Lambda^{A(0)}$ superfields $\Lambda^{A(0)} = \lambda^{A(0)} + \theta F^{A(0)}$, where
$\lambda^{A(0)}$ are right-moving world-sheet fermions and $F^{A(0)}$ auxiliary fields.
The ${\cal N}=1/2$ covariant derivative is defined as
\be
D = \frac{\partial}{\partial\theta} + i \theta \partial\,,
\ee
in complex coordinates, where $\partial = \partial/\partial z = \partial_\tau + i\partial_\sigma$.
The $\Z_2$ action implies that
$(Z_k^{(0)},\Lambda_{1,2}^{(0)},X)$ $\sim$ $(-Z_k^{(0)},-\Lambda_{1,2}^{(0)}, X+\pi R)$.
It will be useful, in what follows, to define world-sheet superfields
\bea
Z_{k} =  Z_{k}^{(0)}\exp(i\eta_k X/R), \ \ \ \bar Z_{k} =\bar Z_{k}^{(0)}
\exp(-i \eta_k X/R)\,, \nn \\
\Lambda_{k}= \Lambda_{k}^{(0)} \exp(i\eta_k X/R), \ \ \
\Lambda_{k}^\star =\Lambda_{k}^{\star (0)} \exp(-i\eta_k X/R)\,,
\label{change-coord}
\eea
where $\eta_1=1$, $\eta_2=-1$,
so that all fields become single-valued along the SS direction $x$.
In this way, integrating over the Grassmanian variable $\theta$
and solving for the auxiliary fields $F_A^*$ and $F^A$ one finds
\be
{\cal L} = g_{\mu\nu} \partial x^\mu \bar\partial x^\nu +
i g_{\mu\nu} \psi^\mu \Big[\bar\partial \psi^\nu +
\Gamma^\nu_{\rho\sigma} \bar\partial x^\rho \psi^\sigma\Big] +
\lambda_A^\star \Big[\partial \lambda^A +  A^A_{x}\partial x^x  \lambda^A  \Big]\,,
\label{2d-2}
\ee
where $x^\mu = (x,z^1,\bar z^1,z^2,\bar z^2)$, $A_x^A=1/R(-i,i,0,\ldots,0)$ is a
discrete $\Z_2$ Wilson line\footnote{This is a non-trivial
Wilson line because, due to the shift, the effective radius of the SS direction is $R/2$,
rather than $R$.}, and $g_{\mu\nu}$ is the metric
\be
ds^2 = g_{\mu\nu}dx^\mu dx^\nu = \Big(1+\frac{|z_1|^2+|z_2|^2}{R^2}\Big)dx^2 +
dz_k d\bar z_k - \frac {i\eta_k}{R} dx (z_k d\bar z_k - \bar z_k d z_k)\,.
\label{g-melvin}
\ee
The metric (\ref{g-melvin}) is a generalization of the 4D Melvin background metric
for a compact space, that is known to correspond to a SS dimensional reduction on a circle of a
higher dimensional flat space \cite{melvin}. Notice that the periodicity conditions for the $z^k$'s
have now changed and become $x$-dependent: if $z^{k(0)}\sim z^{k(0)} +1 \sim z^{k(0)} + i$,
the new coordinates satisfy the periodicity conditions
\be
z^{k}\sim z^{k} + e^{i\eta_k x/R} \sim z^{k} + i e^{i\eta_k x/R}\,.
\label{new-per-cond}
\ee
The $T^4$ torus is now non-trivially fibered along the SS direction $x$.
{}From the construction above, it is clear that such backgrounds are exact classical
solutions to all orders in $\alpha^\prime$, being related by a simple rescaling
to a free SCFT. The latter theory has actually ${\cal N}=(4,0)$ world-sheet SUSY
and implies that the background manifold is an hyper-K\"ahler manifold, admitting three
complex structures. Since we are not going to fully exploit the hyper-K\"ahler
structure of this background, but rather only its property of being complex,
we pick up a particular complex structure $J$. The latter turns out to play
an important role in the following, and thus we explicitly derive its form.
The complex structure $J$ associated to the original freely-acting
orbifold $S^1\times (T^4\times S^1)/{\bf Z}_2$, being globally well-defined despite
the $\Z_2$ shift, can be taken to be the trivial one, as in flat space.
After the change of coordinates (\ref{change-coord}),
$J$ takes the form
\be
J^\mu_{\;\;\nu} = \left(
\matrix{
\;\; 0 \a -1 \a \;\;0 \a \;\;0 \a \;\;0 \a \;\;0 \cr
-1 \a  \;\; 0 \a \;\;0 \a \;\;0 \a \;\;0 \a \;\;0 \cr
 \;\; x_4/R \a \;\; -x_3/R \a \;\;0 \a  \;\;1 \a \;\;0 \a \;\;0  \cr
 \;\; -x_3/R \a - x_4/R \a -1 \a \;\;0 \a \;\;0 \a \;\;0 \cr
 \;\;-x_6/R \a \;\; x_5/R \a \;\;0 \a \;\;0 \a \;\;0 \a \;\;1  \cr
 \;\;x_5/R \a x_6/R \a \;\;0 \a \;\;0 \a -1 \a \;\;0 \cr}
\right)\;,\label{Rtwist}
\ee
where the first two entries in $J$ are taken respectively along the directions of the
two circles where the $\Z_2$ does not ($x_1$ direction) and does act ($x_2\equiv x$ direction),
and we have introduced real coordinates defined as $z_k = x_{2k+1}+i x_{2k+2}$.
It is straightforward to verify that $J$ is actually a complex structure,
with $J^2=-I$ and vanishing associated Nijenhuis tensor.

The analysis can be easily extended to the $\Z_N$ case, acting as
$x\rightarrow x+2\pi R/N$, $z_k\rightarrow e^{2i\pi v_k} z_k$, where $v_k=\frac 1N(-1,1)$.
In this case the periodicity is
$(Z_k^{(0)},\Lambda_{k}^{(0)},X)$ $\sim
(e^{2i\pi\eta_k/N} Z_k^{(0)},e^{2i\pi\eta_k/N}\Lambda_{k}^{(0)}, X+2\pi R/N)$ and can be solved by
the same redefinition (\ref{change-coord}), so that the final geometry (Wilson line included)
is the same.

\subsection{Massless spectrum}

The massless spectrum of this model is easily obtained, being closely related to that
of the heterotic string on the well-known $T^4/\Z_2\times T^2$ orbifold.
In terms of 4D ${\cal N}=2$ SUSY multiplets, we get one gravitational multiplet,
3 $U(1)$ vector multiplets and 4 hypermultiplets from the gravitational sector,
{\em i.e.} from the decomposition of the 10D metric, antisymmetric tensor field
and dilaton. The gauge sector is also straightforward. Focusing on the $SO(32)$ case,
we have one vectormultiplet in the adjoint of the $SO(28)\times SO(4)$ group, that
is the unbroken gauge group in 4D, and one hypermultiplet
in the bifundamental $({\bf 28},{\bf 4})$.

Notice that this spectrum is essentially
a truncation of that of the $SO(32)$ heterotic string on $T^4/\Z_2\times T^2$.\footnote{This is true
only at the massless level.}
In the $T^4/\Z_2\times T^2$ case, we would have obtained all the states as before, but in addition
other states arising from twisted sectors. More precisely, 16 neutral
and 16 charged hypermultiplets (one for each of the 16 fixed points of $T^4/\Z_2$),
the latter in the $({\bf 28},{\bf 4})$ representation of $SO(28)\times SO(4)$,
with ${\bf 4}$ the spinor representation of $SO(4)$.
In presence of the $\Z_2$ shift, the twisted vacuum state carries a non-trivial
winding number and is thus massive. We see, then, that compared to the
$T^4/\Z_2\times T^2$ orbifold case (no shift), many moduli (geometrical and not)
have been lifted, precisely like in presence of fluxes.

\section{The T-dual model}

The metric (\ref{g-melvin}) admits locally an isometry along the SS direction $x$, but
the $x$ direction enters also in the periodicity conditions (\ref{new-per-cond}).
Correspondingly, no isometry is actually allowed and it is not possible to T-dualize
along the $x$ direction\footnote{This
seems to have been overlooked in \cite{K3} where, however, these vacua
have been neglected for other reasons.}.
This can be seen by making use of the world-sheet approach
developed in \cite{Rocek} for the Type II case. In this formalism,
the two vacua related by T-duality arise as two different realizations
of a single world-sheet theory, where the isometry under which one is T-dualizing,
is gauged. We do not find any gauged model that can satisfy
the periodicity conditions (\ref{new-per-cond}).
The impossibility of performing such T-duality transformation in the compact case
is easily understood if one considers a Type IIB model on the orbifold we are considering.
If this T-duality transformation would exist, it would give rise to a
consistent and exact (to all orders in $\alpha^\prime$) 4D vacuum of Type IIB string
theory with NSNS flux turned on.
However, this is not a consistent background since it does not satisfy
the conditions found in \cite{K1} for Type IIB compactifications down to four
dimensions\footnote{On the contrary, its non-compact limit is a suitable and exact solution
of Type II string theory.}.

The same problem arises in the heterotic case. Hence, in order to avoid this
obstruction, we consider the non-compact
limit of the above string vacua, where the model looks like
$R^4\times S^1 \times (C^2\times T^2)/\Z_2$.
In this case, $\partial/\partial x$ defines an isometry and we can perform a $T$-duality
transformation.

The T-duality rules for generic non-flat backgrounds are known
\cite{Buscher}. In the heterotic case we are considering,
they have been derived by \cite{Hassan1} at leading order in a derivative expansion
from a low-energy effective action point of view.
As far as we know, so far there is no satisfactory world-sheet derivation of such
rules\footnote{Ref.\cite{Alva} discusses a $\sigma$-model
approach to T-duality in heterotic theories, but they do not seem to recover the
usual T-duality rules for simple toroidal compactifications, missing some corrections due to
Wilson lines.}. For our purposes, it will be useful to consider directly the
T-dual version of the complex structure $J$. Following \cite{Hassan1,Hassan2},
simple T-duality rules for the inverse metric $g^{-1}$, the complex structure $J$ and the gauge
connection $A$ can be written in terms of a
matrix $Q^\mu_{\;\;\nu}$
(see \cite{Hassan2} for details), so that
\bea
J = Q \, \tilde J \, Q^{-1}\,,\ \ \
g =  Q \, \tilde g \, Q^T \,, \ \ \ \label{T-dual-back}
A^A =  \tilde A^A \, Q^{-1}  \,,
\eea
where we denote by a tilde the original fields, before the T-duality transformation.
The dual dilaton is as usual
\be
\Phi =  \tilde\Phi -\frac 14 \log\Bigg[\frac{{\rm Det}\,\tilde g}{{\rm Det}\, g}\Bigg]\,.
\ee
The background defined by (\ref{T-dual-back}) corresponds
to an heterotic SUSY vacuum on a non-K\"ahler manifold with non-trivial torsion $H$.
Such class of backgrounds have been studied by \cite{Strominger}, where the requirements
of ${\cal N}=1$ SUSY were derived. We use the notation of \cite{polchBook}
that differs from the original introduced in \cite{Strominger}, $\Phi_S$, $H_S$ and $F_S$,
by the rescaling:
\bea
\Phi=-4\Phi_S\,,\,\,\, H=2H_S\,,\,\,\, F=2^{3/2} F_S\,.
\eea
We can summarize the requirements as
\bea
d^\dagger \hat J =  2i (\partial - \bar \partial) \Phi\,, \nn \\
\hat J^{a\bar b} F_{a\bar b} =  0 \,,\label{EQM}  \\
F_{ab} = F_{\bar a \bar b} =  0\,, \nn
\eea
where $a,b,\ldots$ and $\bar a,\bar b,\ldots$ are respectively holomorphic
and anti-holomorphic indices, with respect to the complex structure $J$,
$F$ is the Yang-Mills field strength and $\hat J$ is the
fundamental (1,1)-form obtained by $J$: $\hat J_{\mu \nu}= g_{\mu\rho} J^{\rho}_{\;\;\nu}$.
The torsion $H$ associated to the background is expressed in terms of $\hat J$ as
follows:
\be
H = i(\partial - \bar \partial) \hat J\,,
\label{H-def}
\ee
where the torsion $H$ should satisfy the Bianchi identity
\be
dH = \frac{{\alpha^\prime}}{4} \Big({\rm tr} \, R^2 - {\rm tr} F^2 \Big)\,.
\label{Bianchi}
\ee
The equations of motion (\ref{EQM}), as well as the definition (\ref{H-def})
for $H$ or the Bianchi identity (\ref{Bianchi}) are local expressions valid for any
six-dimensional compactification manifold, and hence should be satisfied also
in the non-compact limit we are considering.
Starting from the explicit form of (\ref{T-dual-back}), it is straightforward, although
laborious, to verify that the equations of motion (\ref{EQM}) are, in fact, exactly
verified. Although in the original model $F=0$, the T-dual field strength is non vanishing
and thus the last two equations in (\ref{EQM}) are satisfied in a non-trivial way.
Notice that is not necessary to go to complex coordinates to verify (\ref{EQM})
or the Bianchi identity (\ref{Bianchi}).

On the other hand, the torsion $H$, as defined in (\ref{H-def}), does not satisfy the
Bianchi identity (\ref{Bianchi}), but only its two-derivative version
\be
dH = -\frac{{\alpha^\prime}}{4} {\rm tr} F^2 \,.
\label{Bianchi-app}
\ee
As mentioned in the introduction, this discrepancy is due to
the T-duality transformation rules we have used \cite{Hassan1}, that neglects
the 4-derivative term ${\rm tr}\,R^2$.
Since higher-order corrections to the T-duality rules are in general
expected for non-trivial backgrounds, we see that no inconsistency arises.
On the contrary, the equations of motion (\ref{EQM}) are satisfied by our background exactly,
and not only at leading order in $\alpha^\prime$.
We think that this
has to do with the higher degree of symmetry we have,
in particular to the fact that our background is actually ${\cal N}=2$ space-time
supersymmetric, with an associated ${\cal N}=(4,0)$ SCFT.
The latter SCFT have been shown to be finite (see {\em e.g.} \cite{Howe}).
This implies that the T-duality
rules for these backgrounds do not get higher order $\alpha^\prime$ corrections and thus lead
to T-dual exact backgrounds. However, as had been already pointed out in \cite{Howe}
and emphasized in \cite{Callan}, the above UV finiteness could be spoiled by one-loop
sigma-model chiral anomalies, whose cancellation requires the well-known modification
to the Bianchi identity for the NSNS field $H$ \cite{Hull:jv}.
This would result in a modification of the T-duality rules for our background that
affects the Bianchi identity for $H$ only. In order to verify the above statements, it
is necessary to compute the corrections to the T-duality rules for heterotic theories,
that are not completely established so far.

\subsection{Geometric description}

Supersymmetric string vacua with non-vanishing fluxes are best classified by the group structures
(or $G$-structures), rather than the holonomy, of the compactification manifold.
Roughly speaking, a $d$-dimensional manifold admits a group structure $G\subseteq SO(d)$
if all tensors (and spinors) can be decomposed globally into representations of $G$.
Classifications of a large class of supersymmetric string and M-theory vacua in terms of
$G$-structures has been derived in \cite{Martelli}. In a 4D heterotic context,
it has been shown in \cite{DA1} how $SU(3)$-structures are particularly useful
in classifying vacua with torsion. The latter can be decomposed into 5 classes, denoted
${\cal W}_i$, $i=1,\ldots,5$, according to their different representations under $SU(3)$.
The equations of motion (\ref{EQM}) and the relation (\ref{H-def}) between the complex structure
$J$ and the torsion $H$ can be rephrased as a constraint on the possible torsion classes of $H$.
One finds that \cite{DA1} ${\cal W}_1$, ${\cal W}_2$ and the combination
$2{\cal W}_4+{\cal W}_5$ must vanish, with ${\cal W}_4$ and ${\cal W}_5$ real and exact.
All the above considerations must hold also in the non-compact limit and thus apply
to our T-dual heterotic configuration.
In what follows, along the lines of \cite{DA1}, we compute ${\cal W}_i$ corresponding
to our particular string vacuum and show that it is actually of the most general form,
where all three classes ${\cal W}_3$, ${\cal W}_4$ and ${\cal W}_5$ are
non-vanishing\footnote{Since our background has ${\cal N}=2$, rather than ${\cal N}=1$, SUSY,
a more refined classification in terms of $SU(2)$-structures should be possible.
We did not find an easy way to do that, and hence we restrict our attention to $SU(3)$-structures.}.

We introduce a basis of vielbeins $e^i$, $i=1,\ldots,6$ (see the Appendix for their explicit
expression) so that the complex structure $J$ reads
\be
J = e^1 \wedge e^2 + e^3 \wedge e^4 + e^5 \wedge e^6 \,.
\ee
It is useful to define a (3,0)-form (with respect to the above defined complex structure)
$\Psi$
\be
\Psi = (e^1 + i e^2) \wedge (e^3 + i e^4) \wedge (e^5 + i e^6) \,.
\ee
The requirement that ${\cal W}_1={\cal W}_2=0$ implies respectively that
$dJ = d\Psi = 0$, as can be easily verified. This is a simple consistency check, since
${\cal W}_1={\cal W}_2=0$ is a necessary and sufficient condition for the manifold
to be complex, a condition that we have already explicitly checked.
The torsion class ${\cal W}_4$ can be directly derived from the dilaton $\Phi$:
\be
{\cal W}_4 = d\Phi = -\frac{1}{f^2} (\sum_{i=3}^6 x_i dx_i)\,,
\ee
where
\be
f=R^2+x_3^2+x_4^2+x_5^2+x_6^2+\alpha^\prime\omega,
\ee
and $\omega$ is a constant related to the Wilson line of the original model. It respectively equals
+1 and 0 for the heterotic and Type II strings.
On the other hand, ${\cal W}_5$ can be computed starting from the real part of $\Psi$
(see \cite{DA1} for details) and satisfies the relation $2{\cal W}_4+{\cal W}_5=0$.
Finally, ${\cal W}_3$ is obtained by taking the (2,1)-form from $dJ-J\wedge {\cal W}_4$ and
it is non-vanishing.

\section{Conclusions}

In this paper, a precise correspondence between the Scherk-Schwarz
symmetry breaking mechanism and fluxes has been given, showing
how T-duality relates them. 
The SS twist under consideration is geometric, namely the space-time
fields are twisted according to their representations under the
Lorentz group in the internal directions. The twist modifies the geometry
of the compact space, that is given by a generalized Melvin geometry.
It can be equivalently seen as a compactification on a flat space
with a constant spin-connection background, the two pictures
being related by a local Lorentz transformation and thus gauge-equivalent
\cite{hos}.

We focused on a particular $\Z_2$ orbifold, given by a twisted four-torus
fibered along a given extra circle, but the construction
can be straightforwardly extended to generic $\Z_N$ orbifolds, with $N>2$.
Interestingly, one can similarly study 3D theories compactified 
on a 7D space given by a six-torus fibered along an extra circle. The latter
is an explicit instance of a smooth 7D manifold with $G_2$ holonomy.
The generalization of our work to a 6D manifold of $SU(3)$ holonomy, in order
to get ${\cal N}=1$ rather than ${\cal N}=2$ 4D SUSY, is instead non-trivial,
reflecting the known difficulties of getting an explicit metric for smooth Calabi-Yau 
manifolds.

By taking the non-compact limit of the above $\Z_2$ model, where a certain
T-duality can be performed, we get a background with non-trivial torsion and 
non-trivial $SU(2)$ structures.  The same T-duality transformation, applied 
to the above 3D model, would give rise to a dual background with non-trivial 
$G_2$ structures.

We have shown how to relate a string vacuum with flux to a simple freely-acting
orbifold. The latter is described by a free SCFT and thus allows a detailed
study of these backgrounds at a full-fledged string level.
Two important issues, however, must be addressed: the corrections to the T-duality
rules for heterotic models, in order to get the corrected Bianchi identity, and
how to extend the construction to compact spaces.

\section*{Acknowledgements}

We thank L. \'Alvarez-Gaum\'e, L.~Bonora, G.~Dall'Agata, E.~Gava, S.~F.~Hassan, D.~Martelli
and K.~S.~Narain for useful discussions.

Work partially supported by the EC through the RTN network
``The quantum structure of space-time and the
geometric nature of fundamental interactions'', contract HPRN-CT-2000-00131.

\appendix

\section{Explicit background}

In this appendix, we report the explicit expressions for the
complex structure $J$, vielbein $e$, gauge connection $A$
and NSNS field strength $H$ of our model, T-dual of a generalized
Melvin (or SS) background.

The final form of the metric is
\bea
\nonumber
ds^2=dx_1^2+\frac{R^2(f-\omega\alpha^\prime)}{f^2} 
dx_2^2 +\frac{2R}{f^2}dx_2(x_4dx_3-x_4dx_3+x_5dx_6-x_6dx_5)+\\
\sum_{i=3}^6 dx_i^2-\frac{f+\omega\alpha^\prime}{f^2}
\left[ (x_4dx_3-x_3dx_4)^2+(x_5dx_6-x_6dx_5)^2-\right.\\\left.
(x_4x_6dx_3dx_5-x_4x_5dx_3dx_6-x_3x_6dx_4dx_5+x_3x_5dx_4dx_6)\right] \nonumber
\eea
where $f$ was defined before as
\bea
f=R^2+x_3^2+x_4^2+x_5^2+x_6^2+\omega\al.
\eea

A possible set of vielbein for this metric, given by the duality, is
\bea
\a\a e_\mu^1=           \left(1,\, 0,   \,0,      \,0,\,0,\,0\right),\nonumber\\
\a\a e_\mu^2=\frac{R}{f}\left(0,\, R,   \,-x_4,   \,x_3,    \, x_6,   \,-x_5\right),\nonumber\\
\a\a e_\mu^3=\frac{1}{f}\left(0,\, Rx_4,\,f-x_4^2,\,x_3x_4, \, x_4x_6,\,-x_4x_5\right),\\
\a\a e_\mu^4=\frac{1}{f}\left(0,\,-Rx_3,\, x_3x_4,\,f-x_3^2,\,-x_3x_6,\, x_3x_5\right),\nonumber\\
\a\a e_\mu^5=\frac{1}{f}\left(0,\,-Rx_6,\, x_4x_6,\,-x_3x_6,\,f-x_6^2,\, x_5x_6\right),\nonumber\\
\a\a e_\mu^6=\frac{1}{f}\left(0,\, Rx_5,\,-x_4x_5,\, x_3x_5,\, x_5x_6,\,f-x_5^2\right).\nonumber
\eea

The $B$-field takes the form

\bea
B=\frac{R}{f}\left(\begin{array}{rrrrrr}
      0   &     0     &  0   &   0   &  0   &   0   \\
      0   &     0     & -x_4 &  x_3  &  x_6 &  -x_5 \\
      0   &    x_4    &  0   &   0   &   0  &   0   \\
      0   &   -x_3    &  0   &   0   &   0  &   0   \\
      0   &   -x_6    &  0   &   0   &   0  &   0   \\
      0   &    x_5    &  0   &   0   &   0  &   0   \\
\end{array}
\right).
\eea

The Wilson Lines are mapped to a non trivial background for the gauge field:
\bea
A_1=-A_2=-\frac{i}{f}\left(
0\,,\,\,\, R \,,\,\,\, - x_4\,, \,\,\,  x_3\,, \,\,\, - x_6\,, \,\,\,  x_5\right).
\eea

Also the complex structure is mapped to a new one, $J$, that is still a complex
structure for the new metric and has vanishing Nijenhuis tensor:
\bea
J=\frac{1}{f}\left(\begin{array}{cccccccccccc}
        0             &&    R^2     &&    -x_4R      &&\hs x_3R      &&\hs x_6 R     &&   -x_5R       \\
-f(1+\omega\al R^{-2})&&     0      &&    x_3 fR^{-1}&& x_4 fR^{-1}  &&-x_5 fR^{-1}  &&   -x_6 fR^{-1}    \\
\hs x_4 fR^{-1}       &&   -x_3 R   &&\hs x_3 x_4    &&    f-x_3^2   &&   -x_3 x_6   &&\hs x_3 x_5    \\
   -x_3 fR^{-1}       &&   -x_4 R   &&   -f+x_4^2    &&   -x_3 x_4   &&   -x_4 x_6   &&\hs x_4 x_5    \\
   -x_6 fR^{-1}       &&\hs x_5 R   &&   -x_4 x_5    &&\hs x_3 x_5   &&\hs x_5 x_6   &&    f-x_5^2    \\
\hs x_5 fR^{-1}       &&\hs x_6 R   &&   -x_4 x_6    &&\hs x_3 x_4   &&   -f+x_6^2   &&  -x_5 x_6
\end{array}
\right).
\eea

The form of $H$ is easily given in components:
\bea
H_{234}=-\frac{2 R}{f^2} (R^2+x_5^2+x_6^2),&&\,\,\,\,
H_{235}=-H_{246}= \frac{2R}{f^2} (x_4x_5+x_3x_6),\nonumber\\
H_{256}=-\frac{2 R}{f^2} (R^2+x_3^2+x_4^2),&&\,\,\,\, H_{236}\,\,=\,\,\,H_{245}= \frac{2R}{f^2} (x_4x_6-x_3x_5),\\\nonumber
H_{1ab}=0,&&\,\,\,\,
H_{\hat a\hat b\hat c}=\frac{2 \omega\al}{f^2}
\sum_{d=3}^6\epsilon_{\hat a\hat b\hat c\hat d} \,\,x_{\hat d},
\eea
where $\hat a$ is an index running only along the directions $\{3,4,5,6\}$ and
$\epsilon_{\hat a\hat b\hat c\hat d}$ is the usual totally antisymmetric tensor of
the 4-D subspace $x_3\dots x_6$, with $\epsilon_{3456}=1$.

The description here is valid also in the type IIB string case, provided that
one puts the Wilson line parameter $\omega$ to zero.

\end{document}